\documentclass[aps,prx,sd,preprint,longbibliography]{revtex4-1}
\usepackage{graphicx}% Include figure files
\usepackage{bm}% bold math
\usepackage{amssymb}
\usepackage{amsmath}
\usepackage{color}

\def\Ray{\mathcal {R} }
\def\mRay{\tilde{\mathcal {R}}}

\newcommand{\change}[1]{\textcolor{black}{#1}}

\draft % marks overfull lines with a black rule on the right

\begin{document}

\title{ Application of Onsager Machlup integral in solving dynamic equations  in non-equilibrium systems} 
%\title{ Variational solution of non-equilibrium dynamic problems - application of Onsager-Machlup integral - } 
\author{Masao Doi}
\affiliation{Center of Soft Matter Physics and its Applications, Beihang University, 37 Xueyuan Road, Beijing 100191, China}

\author{Jiajia Zhou}
\affiliation{Center of Soft Matter Physics and its Applications, Beihang University, 37 Xueyuan Road, Beijing 100191, China}

\author{Yana Di}
\affiliation{LSEC, ICMSEC, Academy of Mathematics and Systems Science, Chinese Academy of Sciences, Beijing 100190, China}

\author{Xianmin Xu}
\affiliation{LSEC, ICMSEC, Academy of Mathematics and Systems Science, Chinese Academy of Sciences, Beijing 100190, China}

\date{\today}

\begin{abstract}
In 1931, Onsager proposed a variational principle which has become the base of many kinetic equations for non-equilibrium systems. 
We have been showing that this principle is useful in obtaining approximate solutions for the kinetic equations, but our previous method has a weakness that it can be justified, strictly speaking, only for small incremental time.  
Here we propose an improved method which does not have this drawback.
The new method utilizes the integral proposed by Onsager and Machlup in 1953, and can
tell us which of the approximate solutions is the best solution without knowing the exact
solution.  
The new method has an advantage that it allows us to determine the steady state in non-equilibrium system by a variational calculus. 
We demonstrate this using three examples, 
(a) simple diffusion problem, 
(b) capillary problem in a tube with corners, and 
(c) free boundary problem in liquid coating, for which the kinetic equations are written in second or fourth order partial differential equations..
\end{abstract}


\maketitle %\maketitle must follow title, authors, abstract and \pacs

%%%%%%%%%%%%%%%%%%%%%%%%%%%%%%%%%%%%%%%%%%%%%%%%%%%%%%%%%%%
\section{Introduction}

In 1931, Onsager published two papers on the dynamics of non-equilibrium systems \cite{Onsager1931,Onsager1931a}. 
He had shown that if the rate of change of the system is written as a linear function of thermodynamic forces, the coefficients must be symmetric and positive definite.  
The symmetry of the coefficients is called reciprocal relation, and has become the base of classical irreversible thermodynamics \cite{deGroot_Mazur, KjelstrupBedeaux,  Oettinger_beyond, Beris}.  

In the same papers, Onsager proposed a variational principle which is a direct  consequence of the reciprocal relation, and states the essence of his theory (the kinetic equation plus the reciprocal relation) in a compact form.
Many kinetic equations that have been used to describe the non-equilibrium phenomena, 
such as Stokes equation in hydrodynamics, Smoluchowskii equation in particle diffusion, 
Nernst-Planck equation in electro-kinetics, Ericksen-Leslie equation in nematic liquid crystals, and etc, can be derived from this principle \cite{Doi2011, DoiSoft}.  
The variational principle has also been shown to be useful to derive time evolution equations for complex systems, where two or more irreversible processes are coupled, or where the process is constrained by geometry, say on curved surfaces or lines. 

We have been showing that the variational principle is useful not only for deriving the time evolution equations, but also useful for obtaining the solutions of the kinetic equations \cite{Doi2015, XuXianmin2016, ZhouJiajia2018}.  
A way of doing this is to assume certain forms for the solutions which involve some 
time dependent parameters, and determine the parameters by the variational principle.  
We have demonstrated the utility of this method for many examples \cite{DiYana2016, ManXingkun2016, ManXingkun2017, ZhouJiajia2018} \change{in fluid and soft matter systems.
The method was also applied to the study of solid-state dewetting \cite{JiangWei2019}.}

On the other hand, our previous method has a weakness that it is only valid, strictly speaking, to predict the state in near future, i.e., to predict the state at time $t+\Delta t$ from the knowledge of the state at time $t$ for infinitesimally small $\Delta t$.  
Although we can use this method repeatedly to predict the state at times $t+2 \Delta t$, $t+3 \Delta t$, $\cdots$, 
the error may increase with time, and we cannot tell which solution is the best among all possible evolution paths.

Here we propose a new variational method which does not have this weakness. 
In the new method, we do not use the original variational principle proposed by Onsager in 1931, but use the integral proposed by Onsager and Machlup in 1953 \cite{Onsager1953}.  
Compared with the previous method, the calculation of the new method is more cumbersome, but it has an advantage that we can directly determine which kinetic paths is the best among various possible kinetic paths.  
Also, it allows us to determine the steady state directly without solving the time evolution equations. 

In this paper, we first explain the general framework of our new variational method
(Sec.~\ref{sec:formulation}). 
Next we demonstrate this method for a few examples, i.e., by solving simple diffusion equation (Sec.~\ref{sec:diffusion}), and by solving two hydrodynamic problems, the liquid wetting in a tube (Sec.~\ref{sec:finger}), and a free boundary problem associated with coating (Sec.~\ref{sec:coating}).  
%We conclude with a short summary (Sec. \ref{sec:summary})

%%%%%%%%%%%%%%%%%%%%%%%%%%%%%%%%%%%%%%%%%%%%%%%%%%%%%%
\section{Onsager principle \label{sec:formulation}}

%===================================================================
\subsection{Onsager principle in its original form }

First we explain the original variational principle of Onsager \cite{Onsager1931, Onsager1931a}. 
Since our target here is the flow and diffusion in soft matter, we shall limit our discussion to isothermal systems where temperature is assumed to be constant.

We consider a non-equilibrium system which is characterized by a set of state variables $x=(x_1, x_2, \cdots ,x_f)$. 
Our objective is to calculate the time dependence of $x(t)$ for a given initial state 
$x(t=0)$. 
(In this paper, we do not consider fluctuations, but fluctuations can be included in the present framework by considering the distribution function of $x$, see Ref.~\cite{Doi2011}.) 
The time evolution of $x$ is obtained by solving Onsager's kinetic equations
\begin{equation}
  \label{eq:kinetic}
  \frac{d x_i}{d t} = -\sum_j  \mu_{ij} \frac{\partial A}{\partial x_j} ,  
\end{equation}
where $A(x)$ is the free energy of the system and $\mu_{ij}(x)$ are kinetic coefficients, both are functions of $x$.  
Onsager showed that if the kinetic equation is written in the form of Eq.~(\ref{eq:kinetic}), $\mu_{ij}(x)$ must be positive definite and symmetric.
\begin{eqnarray}
  && \sum_{ij} \mu_{ij} \dot x_i \dot x_j \ge 0 \quad \mbox{for arbitrary $\dot x_i$}, \\
  &&  \mu_{ij} =   \mu_{ji}  .    \label{eqn:reciprocal}
  \label{eq:reciprocal}
\end{eqnarray}
Equation (\ref{eqn:reciprocal}) is called Onsager's reciprocal relation.

The kinetic equation (\ref{eq:kinetic}) can be written in the form of force balance equation.  
Let $\zeta_{ij}$ be the inverse matrix  of $\mu_{ij}$ ($\sum_{k}  \zeta_{ik} \mu_{kj} = \delta_{ij}$). 
Equation (\ref{eq:kinetic}) is then written as
\begin{equation}
  \label{eq:force}
   - \frac{\partial A}{\partial x_i} - \sum_j  \zeta_{ij} \frac{d x_j}{d t} =0 .
\end{equation}
The first term represents the thermodynamic force that drives the system to
the state of minimum free energy, and the second term represents the frictional force
which resists against this change. 

The force balance equation (\ref{eq:force}) can be cast in a form of
minimum principle.
Consider the following quadratic function of the rate of change of the state
$\dot x = (\dot x_1, \dot x_2, \cdots, \dot x_f ) $:
\begin{equation}
  \label{eq:Ray}
  \Ray (\dot x; x) = \frac{1}{2} \sum_{i,j}  \zeta_{ij} \dot x_i \dot x_j 
  + \sum_{i}  \frac{\partial A} {\partial x_i}  \dot  x_i .
\end{equation}
The kinetic equation (\ref{eq:force}) is equivalent to the condition that $\Ray$ is minimum with respect to $\dot x$, $\partial \Ray / \partial \dot x_i = 0$, i.e., the time evolution of non-equilibrium system is determined by the minimum principle of
$\Ray (\dot x; x)$.  
This is called Onsager's variational principle, or simply the Onsager principle.

The function $\Ray(\dot x; x)$ is called Rayleighian.
The first term is called the dissipation function
\begin{equation}
  \label{eq:Phi}
  \Phi  = \frac{1}{2} \sum_{i,j}  \zeta_{ij} \dot x_i \dot x_j ,
\end{equation}
and the second term is called the free energy change rate
\begin{equation}
  \label{eq:Adot}
  \dot A = \sum_i  \frac{\partial A}{\partial x_j} \dot  x_i .
\end{equation}
The dissipation function is a quadratic function of $\dot x$, while the free energy change
rate is a linear function of $\dot x$.

%===================================================================
\subsection{Onsager principle as a tool of approximation }

Onsager's variational principle can be used to solve problems by a variational method.
To do this, we assume that $x_i(t)$ ($i=1,2,...$) is written as a certain function of some parameter set $\alpha=(\alpha_1,\alpha_2, \cdots)$, i.e., $x_i(t)$ is written as 
$x_i(t)=x_i(\alpha_1(t),\alpha_2(t), \cdots)$.   
Then $\dot x_i$ is written as
\begin{equation}
  \dot x_i = \sum_{p}  \frac{\partial x_i }{\partial \alpha_p} \dot \alpha_p ,
                           \label{eqn:5a}
\end{equation}
and the Rayleighian is written as
\begin{equation}
  \Ray  = 
     \sum_{pq} \left[ \sum_{ij} \zeta_{ij} 
        \frac{\partial x_i} {\partial \alpha_p} \frac{\partial x_j} {\partial \alpha_q} \right ]
                   \dot \alpha_p\dot \alpha_q  
          + \sum_p \left[  \sum_{i} \frac{\partial A}{\partial x_i} \frac{\partial x_i} {\partial \alpha_p} \right ]
                    \dot  \alpha_p .
                                 \label{eqn:5}
\end{equation}
The time evolution of the parameter $\alpha(t)$ is determined by minimizing $\Ray$ with respect to $\dot \alpha$.  
This method is useful when we have an idea for the kinetic path and can write down the 
functions $x_i(\alpha)$. 
It has been applied to many problems in soft matter \cite{DiYana2016,  ManXingkun2016, XuXianmin2016, ManXingkun2017, ZhouJiajia2017, ZhouJiajia2017a, ZhouJiajia2018}.

%===================================================================
\subsection{Onsager principle in a modified form}

The above variational principle determines the state at $t+\Delta t$ from the knowledge of the state at time $t$.  
It says that among all possible states allowed for the system to be at time $t+\Delta t$, the state chosen by nature is given by the state which minimizes $\Ray ( (x(t+\Delta t)- x(t))/\Delta t; x(t))$.  
This principle can be used to conduct approximate calculation. 
Suppose we have many candidates for the state at time $t+\Delta t$, then we can tell which state is the best candidate: the state which gives the smallest value of the Rayleighian is the best.   

This variational principle is a local principle: it can predict the state in
near future, but cannot predict the state in far future.  
Suppose we have two kinetic paths, starting from the same initial state $x(0)$ and ending at different states  $x(t)$ and $x'(t)$, then we cannot tell which path is better:  we can tell which path is better if $t$ is close to zero, but cannot tell if $t$ is far away from zero. 

This problem can be resolved if we use a slightly modified definition for Rayleighian
\begin{equation}
   \mRay(\dot x; x) = \Ray(\dot x; x) - \Ray_{\rm min}(x),
        \label{eqn:7}
\end{equation}
where $\Ray_{\rm min}(x)$ is the minimum value of $\Ray(\dot x; x)$ in the space of $\dot x$.  
The minimum is given by the velocity
\begin{equation}
   \dot{x}_i^*(x) = - \sum_j \mu_{ij} \frac{\partial A}{\partial x_j}  .        \label{eqn:8}
\end{equation}
Note that $\dot{x}_i^*$ is the actual velocity of the system at state $x$, and depends only on the state variable $x$.

With the use of Eq.~(\ref{eqn:8}),  it is easy to show that $\mRay(\dot x; x)$ is written as 
\begin{equation}
   \mRay(\dot x; x)= \frac{1}{2} \sum_{ij}  \zeta_{ij} (\dot x_i -\dot x_i^*)(\dot x_j -\dot x_j^*) .
        \label{eqn:9}
\end{equation}
Since $\zeta_{ij}$ is positive definite, $ \mRay(\dot x; x)$ is larger than or equal to zero.  

Now consider the following integral, which is a functional of certain kinetic path $x(t)$
\begin{equation}
  \label{eq:OM_integral}
  \mathcal{O} [x(t)] =   \int_0^t dt' \, \mRay(\dot x(t'); x(t'))
  =  \frac{1}{2} \int_0^t dt' 
  \sum_{ij}  \zeta_{ij}(x(t')) \big[\dot x_i(t') -\dot x_i^*(t') \big]
  \big[\dot x_j(t') -\dot x_j^*(t')\big] .
\end{equation}
The integral is positive definite and is equal to zero only when $x(t)$ is equal to the 
actual kinetic path. 
Hence the variational principle can be stated simply that nature chooses the path which minimizes the functional $\mathcal{O}[x(t)]$. 

The integral of Eq.~(\ref{eq:OM_integral}) was first introduced by Onsager and Machlup 
in their discussion on the fluctuation of kinetic paths described by linear Langevin equation \cite{Onsager1953}.  
They have shown that the probability of finding the kinetic path $x(t)$ is 
proportional to $\exp \left[ -\mathcal{O}[x(t)]/\change{2}k_BT \right]$. 
It is easy to show that  their theory can be extended to the general case where both $A$ and $\zeta_{ij}$ are functions of $x$.  
Indeed in some literature \cite{QianTiezheng2006}, this form has been referred to as 
Onsager's variational principle. 
We shall call this variational principle Onsager Machlup principle, and call the integral of Eq.~(\ref{eq:OM_integral}) Onsager Machlup integral.

%=======================================================================
\subsection{Variational calculus using Onsager Machlup integral }

The Onsager Machlup variational principle can be used to obtain the best guess for the kinetic path of the system.  
We consider certain kinetic path which involves a parameter set $\alpha=(\alpha_1, \alpha_2, \cdots)$.
The best guess for the actual path is the path which gives the smallest value of the
Onsager Machlup integral.  
In the following sections, we shall show actual calculation of this method.  
Here we show a few tips that will help such calculations.

The Rayleighian that appears in the Onsager Machlup integral has two velocities, $\dot x_i$ and $\dot x_i^*$.  
For given kinetic path $x(t)$, they are calculated by
\begin{equation}
  \label{eqn:11}
  \dot x_i(t) = \frac{d x_i(t)}{dt}   , \qquad   
  \dot x_i^*(t)= - \left. \sum_j (\zeta^{-1})_{ij} \frac{\partial A(x)}{\partial x_j} \right|_{x=x(t)} .
\end{equation}
Here $\dot x_i$ is defined by the time derivative of the state variables $x(t)$, 
while $\dot x_i^*$ is defined by the thermodynamic force at state $x$.
The Onsager Machlup integral becomes minimum when $\dot x_i $ is equal to $\dot x_i^*$.

At first sight, such variational principle may look useless since to calculate $\mathcal{O}[x(t)]$, 
we need to know $\dot{x}_i^*$, but this is the quantity difficult to calculate; 
if we can calculate $\dot{x}_{\change{i}}^*$, we can directly use it to 
solve the time evolution equation, $d x_i/dt = \dot  x_i^*$ and there is no need to
use such variational principle.  

In fact, the variational principle is useful since we have a freedom to choose the kinetic path $x(t)$.  
We need not calculate $\dot x_i^*$ for a general state; we need to calculate $\dot x_i^*$ only for the state we have chosen.  
By proper choice of the kinetic path, $\dot x_i^*$ may be calculated.  

For example, consider the problem of deformation of a droplet in emulsions under flow \cite{Doi1983}. 
It is difficult to calculate the shape change of the droplet for general shapes, but it can be calculated for special cases where the droplet takes simple shapes such as ellipsoids. 
Therefore, if the droplet keeps a form which can be approximated by ellipsoids, the variational principle can be used to describe the deformation of the droplet.  
We shall see this more clearly in the examples given later.

%======================================================
\subsection{Note on the dissipation function}

So far we have been discussing a simple case where the dissipation function is given explicitly as a function of $\dot x$, as in Eq.~(\ref{eq:Phi}).
In many situations, however, the dissipation function is not given in this form.
Quite often, the dissipation function is written as a quadratic function of other
velocity variables $\dot y_a $ $(a=1,2,...f')$ in an extended space ($f'>f$)
\begin{equation}
  \label{eq:tPhi}
  \tilde \Phi  = \frac{1}{2} \sum_{ab}  \tilde\zeta_{ab} \dot y_a \dot y_b 
  % \label{eqn:13}
\end{equation}
subject to constraints that $\dot x_i$ and $\dot y_a$ are linearly related to each other,
\begin{equation}
  \label{eq:constraint}
   \dot x_i  =  \sum_{a=1}^{f'}  c_{ia}(x) \dot y_a ,
   %\label{eqn:12}
\end{equation}
where the coefficients $c_{ia}$ are some functions of $x$.
Example of such situation is given in the next section.

The dissipation function $\Phi$ in Eq.~(\ref{eq:Phi}) is the minimum of 
$\tilde{\Phi}$ subject to the constraint of Eq. (\ref{eq:constraint}). 
Hence one can state the Onsager principle in the extended space spanned by $\dot y_a (a=1,2,...f')$ \cite{Doi2011}, where the Rayleighian is given by 
\begin{equation}
  \Ray^{(y)}(\dot y; x)  = 
     \frac{1}{2}\sum_{a,b}  \tilde\zeta_{ab} \dot y_a \dot y_b  
                  + \sum_{i,a}  \frac{\partial  A}{\partial x_i} c_{ia} \dot y_a .
                                 \label{eqn:14}
\end{equation}

It is easy to show that the modified Rayleighian in the extended space is given by
\begin{equation}
  \mRay^{(y)}(\dot y; x)  = 
     \frac{1}{2} \sum_{a,b}  \tilde \zeta_{ab} (\dot y_a  - \dot y_a^*)(\dot y_b  - \dot y_b^*)
                                 \label{eqn:15}
\end{equation}
where $\dot y_a^*$ is the velocity which minimizes $\Ray^{(y)}$. 
The Onsager Machlup integral is also given by
\begin{equation}  
  \mathcal{O} [y(t)]   =  \frac{1}{2} \int_0^t dt' 
          \sum_{a,b}  \tilde \zeta_{ab} (\dot y_a  - \dot y_a^*)(\dot y_b  - \dot y_b^*)
         \label{eq:OM_integral1}
\end{equation}

Equations (\ref{eqn:15}) and  (\ref{eq:OM_integral1}) includes two velocities, $\dot y_a$ and $\dot y_a^*$.
The former is determined by the assumed kinetic path $x(\alpha(t))$, while the latter is determined by the thermodynamic forces acting at state $x(\alpha(t))$.  
If the kinetic equations are solved rigorously, they agree with each other, and  $\mRay^{(y)}$ is equal to zero.  
In our approximate treatment, we determine the parameter $\alpha$ so that  $\mRay^{(y)}$ becomes minimum.   
This is the general strategy that we will use in the subsequent examples.

%%%%%%%%%%%%%%%%%%%%%%%%%%%%%%%%%%%%%%%%%%%%%%%%%%%%%%%%%%%%%%
\section{Diffusion \label{sec:diffusion}}

%================================
\subsection{Diffusion equation}

As the first example, we consider the diffusion of particles.
The state variable in this problem is $n(\bm{r})$, the number density of 
particles at position $\bm{r}$, and we consider the time evolution of this function $n(\bm{r};t)$.
The ``velocity'' of the state variable is $\dot n= \partial n/\partial t$, but it is not possible to write down the dissipation function in terms of $\dot n$.  
This is because in diffusion, $\dot n$ must satisfy the conservation equation
\begin{equation}
  \label{eqd:diff1}
  \dot{n} =  - \nabla \cdot \bm{j}
\end{equation}
where $\bm{j}(\bm{r})$ be the flux of the particles.  Equation (\ref{eqd:diff1}) represents the
kinetic constraints that $n$ can change only by diffusion, and that there is no process
(such as chemical reaction) which creates or annihilates the particles.  
The variable $\bm{j}$, introduced to represent such constraints,  is an example of the 
variables $\dot y_a$  given in the previous section.  
The Rayleighian for the diffusion problem can be constructed using $\bm{j}$.

The free energy of the system is written as
\begin{equation}
  \label{eqd:diff2}
  A= \int d \bm{r} f(n(\bm{r}))
\end{equation}
where $f(n)$ is the free energy density of the particles.  $\dot A$ is then calculated by Eqs.~(\ref{eqd:diff1}) and (\ref{eqd:diff2}) as
\begin{equation}
  \label{eqd:diff3}
  \dot A= \int d \bm{r} \frac{\partial f}{\partial n} \dot n     
            = - \int d \bm{r} \frac{\partial f}{\partial n} \nabla \cdot \bm{j}   
            =  \int d \bm{r} \bm{j} \cdot \nabla \left [\frac{\partial f}{\partial n} \right ]   
            =  \int d \bm{r} (\bm{j} \cdot \nabla n)  \frac{\partial^2 f}{\partial n^2}  
\end{equation}

The dissipation function $\Phi$ is proportional to the square of the particle velocity
$ \bm{v}_p =\bm{j}/n$, and can be written as
\begin{equation}
  \label{eqd:diff4}
  \Phi = \frac{1}{2} \int d \bm{r}  n \zeta  \bm{v}_p^2 = \frac{1}{2} \int d \bm{r}  \frac{1}{n} \zeta  \bm{j}^2  
\end{equation}
where $\zeta$ is the friction constant of one particle. 

From Eqs.~(\ref{eqd:diff3}) and (\ref{eqd:diff4}), the Rayleighian is given by  
\begin{equation}
  \label{eqd:diff5}
  \Ray =  \frac{1}{2} \int d \bm{r}  \frac{1}{n} \zeta  \bm{j}^2
        + \int d \bm{r} (\bm{j} \cdot \nabla n ) \frac{\partial^2 f}{\partial n^2}   
\end{equation}
The minimum of this is given by 
\begin{equation}
  \bm j^* = - \frac{n}{\zeta} \frac{\partial^2 f}{\partial n^2} \, \nabla n.
\end{equation} 
Therefore the time evolution equation for $n$ is given by
\begin{equation}
  \label{eqd:diffeq}
  \frac{\partial n}{\partial t} = \nabla \cdot \left[ D(n) \nabla n \right]
\end{equation}
where $D= (n/ \zeta) \partial^2 f/ \partial n^2$.  
Equation (\ref{eqd:diffeq}) is the standard diffusion equation \cite{DoiSoft}.

%====================================
\subsection{Onsager principle as an approximation tool}

We now demonstrate how to use Onsager principle to obtain approximate solutions.
For simplicity, we consider the diffusion in 1-dimension system and in dilute solution.
The conservation equation (\ref{eqd:diff1}) is written as
\begin{equation}
  \label{eqn:D1}
  \dot n = - \frac{\partial j}{\partial x} .
\end{equation}
The free energy of the system is given by
\begin{equation}
  \label{eqn:D2}
  A =  k_B T  \int d x \, n(x) \ln n(x) ,
\end{equation}
and the Rayleighian is given by [see Eq.~(\ref{eqd:diff5})]
\begin{equation}
  \label{eqn:D3}
  \Ray =  \int dx \left[ \frac{1}{2}  \frac{1}{n}  \zeta j^2 
                                 + k_B T  \frac{j}{n} \frac{\partial n}{\partial x} 
                           \right] .
\end{equation}
This becomes minimum when $j$ is equal to
\begin{equation}
  \label{eqn:D3a}
  j^* = - D \frac{\partial n}{\partial x} 
\end{equation}
where  $D= k_BT/ \zeta$.
Equation (\ref{eqn:D3a}) and the conservation equation (\ref{eqn:D1}) give the 1-dimension diffusion equation
\begin{equation}
  \label{eqd:diff6}
  \frac{\partial n}{\partial t} = D \frac{\partial ^2 n}{\partial x^2}   .   
\end{equation}

The exact solution of this equation for the initial condition $n(x,0)=N_0\delta(x)$ is given by
\begin{equation}
  \label{eqd:exact}
  n(x;t) = \frac{N_0 }{\sqrt{4\pi D t}} \exp \left( - \frac{x^2}{4 D t} \right).
\end{equation}

To use the variational principle, we assume that $n(x,t)$ is approximated by the following polynomial function
\begin{equation}
  \label{eqd:trial}
  n(x,t) = N \left[ 1 - \left( \frac{|x|}{a(t)} \right) ^m \right], \quad
  N=N_0 \frac{m+1}{2m} \frac{1}{a},
\end{equation}
where $m$ is a positive integer, $a(t)$ is a parameter characterizing the particle spreading, and $N$ is a constant which has been determined by the condition 
$\int_{-\infty}^{\infty} dx n(x,t) = N_0$.

We shall derive the kinetic equations for $a(t)$ using Onsager principle.
From Eqs.~(\ref{eqn:D2}) and (\ref{eqd:trial}), the free energy is calculated as
\begin{eqnarray}
  A &=& k_B T \int_{-a}^{a}dx  N \left[ 1 - \left( \frac{|x|}{a} \right) ^m \right] 
                         \ln N \left[ 1 - \left( \frac{|x|}{a} \right) ^m \right]   \nonumber \\
     &=&  k_B T \int_{-1}^{1}dz \, a N (1-|z|^m) \ln [ N (1-|z|^m)  ] \nonumber \\
  \label{eqd:A}
     &=& - k_B T N_0 \ln a + \{ \mbox{terms independent of $a$} \}
\end{eqnarray}
where $z=|x|/a$. Hence the free energy change rate is calculated as
\begin{equation}
  \label{eqd:Adot}
  \dot{A} = -k_B T N_0 \frac{\dot{a}}{a} .
\end{equation}

On the other hand, the flux $j(x)$ is calculated by integrating Eq.~(\ref{eqn:D1}) with respect to $x$,
\begin{equation}
  \label{eqd:jx}
  j(x) = -  \int_0^{x} d x' \, \dot n(x') 
    = N_0 \frac{m+1}{2m} \frac{\dot{a}}{a} \left[ \frac{x}{a} - 
      \left(\frac{x}{a} \right)^{m+1} \right]. 
\end{equation}
The dissipation function is then calculated as
\begin{equation}
  \label{eqd:Phi}
  \Phi = \frac{1}{2} \int_{-a}^{a} d x \, \zeta \frac{j^2}{n} 
       = \frac{1}{2} \zeta N_0 \frac{m+1}{3(m+3)} \dot{a}^2 . 
\end{equation}

The Rayleighian is the summation of Eqs.~(\ref{eqd:Adot}) and (\ref{eqd:Phi})
\begin{equation}
  \Ray   
  = \frac{1}{2} \zeta N_0 \frac{m+1}{3(m+3)} \dot{a}^2 - k_B T N_0 \frac{\dot{a}}{a} .
\end{equation}
Minimizing this with respect to $\dot a$, we have 
\begin{equation}
  a \dot{a} = \frac{3(m+3)}{m+1} \frac{k_BT}{\zeta} = \frac{3(m+3)}{m+1} D .
  \label{eqd:kinetics}
\end{equation}
The solution of this equation for the initial condition $a(0)=0$ is
\begin{equation}  \label{eqn:d20}
   a = \left[ \frac{6(m+3)}{m+1} Dt \right]^{1/2} .
\end{equation}
It is seen that the spreading of the particle concentration follows the scaling $a \sim \sqrt{t}$ for all $m$.

%===========================================================
\begin{figure}[htbp]
  \includegraphics[width=0.6\columnwidth]{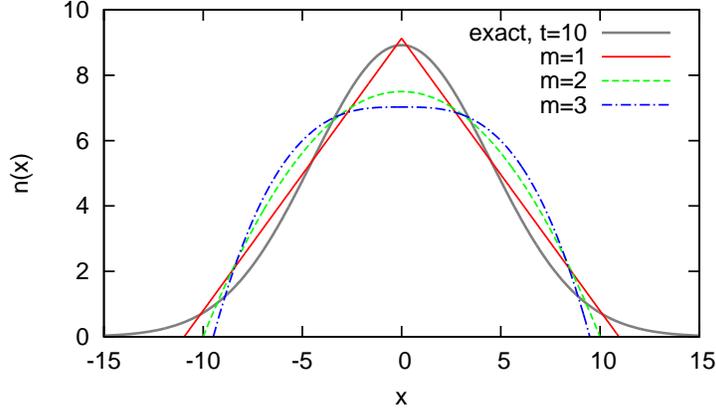}
  \caption{Comparison between the exact solution (\ref{eqd:exact}) and approximated solution (\ref{eqd:kinetics}) for $m=1,2,3$. The parameters are $N_0=100$ and $D=1$. The particle concentrations are shown for $t=10$. }
  \label{fig:diffusion}
\end{figure}
%===========================================================

Figure 1 shows the comparison of the approximate solution $n(x,t)$ 
calculated by Eqs.~(\ref{eqd:trial}) and (\ref{eqd:kinetics}) and the exact solution
(\ref{eqd:exact}).   
From this figure, we can see that the curve of $m=1$ seems to be the best solution 
(the solution closest to the exact solution).  
However, if we do not know the exact solution, we cannot tell which solution is the best.  
This question can be answered if we use the Onsager Machlup integral.

%============================
\subsection{Onsager Machlup integral}

We now analyze the problem using the modified Rayleighian.
In the present problem, the modified Rayleighian is 
obtained by Eqs.~(\ref{eqn:15}) and (\ref{eqd:diff4}) as
\begin{equation}
  \mRay = \frac{1}{2} \int_{-a}^{a} d x \, \frac{\zeta}{n} (j-j^*)^2  \\
\end{equation}
where $j$ is the flux determined by the assumed time dependence of $n(x;t)$, i.e., the
flux given by Eq.~(\ref{eqd:jx})
\begin{equation}
  j(x) = -  \int_0^{x} d x' \, \dot n(x') 
    = N_0 \frac{m+1}{2m} \frac{\dot{a}}{a} \left[ \frac{x}{a} - 
      \left(\frac{x}{a} \right)^{m+1} \right], \qquad \mbox{for} \quad x>0, 
\end{equation}
and $j^*$ is the exact flux at state $n(x,t)$, which is given by
\begin{equation}
  j^*(x) = - D \frac{\partial n}{\partial x} = \frac{DN_0}{a^2} \frac{m+1}{2} \left(\frac{x}{a} \right)^{m-1}  \qquad \mbox{for} \quad x>0.
\end{equation}
The modified Rayleighian is then calculated as
\begin{equation}
  \label{eqd:mRay0}
  \tilde{\mathcal{R}} = \zeta N_0 \left[ \frac{m+1}{6(m+3)} \dot{a}^2 - D \frac{\dot{a}}{a} 
    + \frac{D^2}{2} m (m+1) \left( \int_0^1 d z \frac{z^{2m-2}}{1-z^m} \right) \frac{1}{a^2} \right].
\end{equation}
The evolution is obtained by $\partial \tilde{\mathcal{R}} / \partial \dot{a}=0$, which leads to the same equation as Eq. (\ref{eqd:kinetics}).

Substituting Eq.~(\ref{eqd:kinetics}) into the modified Rayleighian (\ref{eqd:mRay0})
\begin{equation}
  \tilde{\mathcal{R}} = \zeta D N_0   
  \left[ - \frac{1}{4} + \frac{m(m+1)^2}{12(m+3)} 
    \left( \int_0^1 d z \frac{z^{2m-2}}{1-z^m} \right) \right] t^{-1}.
\end{equation}
The integration of the last term diverges as $z \rightarrow 1$. 
We performed the integration by setting the upper bound to $1-\epsilon$. 
For $m=1,2,3$, the results are 
\begin{eqnarray}
  m=1, &\quad& \mRay = \zeta D N_0 \left( -0.250 - 0.083 \,\ln \epsilon \right) t^{-1} \nonumber \\
  m=2, &\quad& \mRay = \zeta D N_0 \left( -\change{0.446} - 0.150 \,\ln \epsilon \right) t^{-1} \nonumber \\
  m=3, &\quad& \mRay = \zeta D N_0 \left( -\change{0.864} - 0.222 \,\ln \epsilon \right) t^{-1} \nonumber 
\end{eqnarray}

If we compare the front factor of the divergent terms, the trial function $m=1$ gives the smallest value.  
Also for a typical small value of $\epsilon=10^{-3}$, the terms in the brackets are evaluated to be 0.3233, \change{0.5902, 0.6695} for $m=1,2,3$, respectively.  
In both cases, the trial function of $m=1$ gives the smallest value, and we can conclude that the function of $m=1$ is the best.
This is consistent with the comparison shown in Fig.~\ref{fig:diffusion}.

%%%%%%%%%%%%%%%%%%%%%%%%%%%%%%%%%%%%%%%%%%%%%%%%%%%%%%%%%%%%%%

\section{Finger flow in a square tube  \label{sec:finger}}
\begin{figure}[htbp]
  \includegraphics[width=0.6\columnwidth]{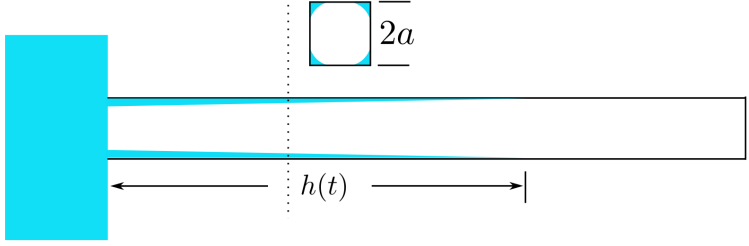}
  \caption{Finger flow in a closed square tube.}
  \label{fig:sketch}
\end{figure}

As the second example, we consider the liquid wetting in a square tube (see Fig.~\ref{fig:sketch}).  
The liquid is contained in the left reservoir and connected to a tube having a square cross section. 
The end of the tube is closed, so the meniscus of the bulk liquid cannot move, but the liquid can advance along the corner of the tube forming a ``finger'' to reduce the surface energy. 
The tube is assumed to be placed horizontally, so gravity does not play a role in this problem.  

The wetting dynamics of such situation was first studied by Dong and Chatzis \cite{Dong1995} \change{using a variational method \cite{Mayer1983}. (See also Ref. \cite{WeiYueXing2011} for the application.)} 
Here we use the same model as theirs, and solve the problem using Onsager principle.
The governing equation of this problem turns out to be the same as the non-linear diffusion equation, and we shall show that the equation can be solved in 
good approximation by using the Onsager Machlup integral.

%=================================================
\subsection{Dynamical equation}
We take $x$ coordinate along the tube axis. We define a dimensionless quantity called 
saturation $s(x;t)$ , which is the 
fraction of the liquid area in the cross section at position $x$ and time $t$. 
We focus on the motion of the finger part, and ignore the motion of the bulk part. 
We set the bottom of the finger at $x=0$, and assume, as in the previous 
works \cite{Dong1995, YuTian2018}, that $s(x,t)$ takes a constant value $s^*$ at $x=0$ 

The time evolution equation for $s(x,t)$ can be derived from the Onsager principle. Since the gravity is ignorable, the free energy of the system is given by the surface energy of the liquid.  
The surface area of the liquid in the region between $x$ and $x+dx$ is proportional to $a\sqrt{s} dx$.  
Hence the free energy of the system can be written as
\begin{equation}
  \label{eqn:tube1}
  A =  - \alpha_1 a \gamma \int_0^h \sqrt{s} \, d x .
\end{equation}
where $\alpha_1$ is a certain numerical constant ($\alpha_1 = 8\sqrt{1-\pi/4}$, 
see Ref.~\cite{YuTian2018}).

The saturation $s(x;t)$ must satisfy the conservation equation
\begin{equation}
  \label{eqn:tube2}
  \frac{\partial s}{\partial t} = - \frac{\partial j}{\partial x}.
\end{equation}
The dissipation function is a quadratic function of $j(x)$.  By
using the lubrication approximation, one can show that the dissipation function is written as
\begin{equation}
  \label{eqn:tube3}
  \Phi = \frac{1}{2} \alpha_2 \eta \int_0^h \frac{j^2}{ s^2} \, d x .
\end{equation}
where $\eta$ is the liquid viscosity and $\alpha_2$ is another 
numerical constant. The hydrodynamic 
calculation \cite{Ransohoff1988, YuTian2018} indicates that  $\alpha_2 \simeq 90$.

The Rayleighian is given by Eq.~(\ref{eqn:tube1}) and Eq.~(\ref{eqn:tube3}), and 
the same procedure as in the previous section gives the following expression for the
fluid flux
\begin{equation}
  \label{eqf:Qfull}
  j^* = - D s^{1/2} \frac{\partial s}{\partial x} 
\end{equation}
where $D$ is defined by 
\begin{equation}
  \label{eqf:pde1}
  D = \frac{\alpha_1}{2\alpha_2} \frac{a \gamma}{\eta},
\end{equation}
which has the same dimension as the diffusion constant.

The time evolution equation is given by
\begin{equation}
  \label{eqf:pde}
  \frac{\partial s}{\partial t} = D \frac{\partial}{\partial x} \left( s^{1/2} \frac{\partial s}{\partial x} \right).
\end{equation}
This is the same equation as that derived by Dong et al.\cite{Dong1995}. 

The boundary conditions and the initial conditions for $s(x,t)$ are
\begin{eqnarray}
  \label{eqf:bc}
  t>0: \quad s(x=0)=s^*, && \quad s(x\rightarrow \infty) = 0,\\
  \label{eqf:ic}
  t=0: \quad s(x=0)=s^*, && \quad s(x>0)=0.
\end{eqnarray}
Dong et al. solved Eq.~(\ref{eqf:pde}) under these conditions, and showed that the liquid front $h(t)$ (see Fig. \ref{fig:sketch})  increases with time as  
\begin{equation}
  h(t) = \sqrt{ \frac{ a \gamma}{\eta} } H t^{1/2}, \quad H \simeq 0.1281.
\end{equation}
In the following we shall obtain $h(t)$ using Onsager Machlup integral.

%%%%%%%%%%%%%%%%%%%%%%%%%%%%%%%%%%%%%%%%%%%%%%%%%%%%%%%%%%%%
\subsection{Onsager Machlup integral}

We assume the profile can be written in the form 
\begin{equation}
  \label{eqf:on}
  s(x;t) = s^* \left[ 1- \frac{x}{h(t)} \right]^n ,
\end{equation}
This function is chosen to satisfy the boundary condition $s(0,t)=s^*$ and the condition $s(h(t);t)=0$. 
The function includes two parameters, $h(t)$ and $n$, which we shall determine using the Onsager principle.

The flux at position $x$ is given by integrating Eq.~(\ref{eqn:tube2}) and using the
condition that $j$ becomes 0 at $x=h(t)$
\begin{equation}
  \label{eqf:Qn}
  j(x) = \int_x^h \dot s(x') d x' = \dot{h} \int_z^1 \frac{\partial s}{\partial z'} (-z') d z' = \dot{h} F(z).
\end{equation} 
where $z=x/h(t)$, and the function $F(z)$ is given by
\begin{equation}
  F(z) = \int_z^1 \frac{d s}{d z} (-z) d z =  \frac{s^*}{n+1} (1-z)^n (1+nz) .
\end{equation}

On the other hand, the flux $j^*$ is given by Eq. (\ref{eqf:Qfull})
\begin{equation}
  \label{eqf:Qfull2}
  j^* = - D s^{1/2} \frac{\partial s}{\partial x} = - D s^{1/2} \frac{\partial s}{\partial z} \frac{1}{h}.
\end{equation}

The modified Rayleighian is given by
\begin{equation}
  \mRay = \frac{1}{2} \alpha_2 \eta \int_0^h  \frac{(j - j^*)^2}{ s^2}  d x.
\end{equation}
Using Eqs. (\ref{eqf:Qn}) and (\ref{eqf:Qfull2}), we can get
\begin{eqnarray}
  \mRay &=& 
    \frac{\alpha_2 \eta}{2} \int_0^1 \frac{1}{ s^2} 
       \left[ F(z) \dot{h} + D s^{1/2} \frac{\partial s}{\partial z} \frac{1}{h} \right]^2 d z \nonumber \\
  &=& \frac{\alpha_2 \eta}{2} \Bigg[ \left( \int_0^1 \frac{F^2(z)}{s^2(z)} d z \right) h \dot{h}^2
       + 2 D \left( \int_0^1 \frac{F(z)}{s^{3/2}(z)} \frac{\partial s}{\partial z} d z \right) \dot{h} 
       + D^2 \left( \int_0^1 \frac{1}{s(z)} \left( \frac{\partial s}{\partial z} \right)^2  d z \right) \frac{1}{h} \Bigg] \nonumber \\
  \label{eqf:mRay}
  &=& \frac{ \alpha_2 \eta }{2} \left[ \frac{n^2+3n+3}{3(n+1)^2} h \dot{h}^2 
     - 2 D \sqrt{s^*} \frac{ 2(3n+2)}{(n+1)(n+2)} \dot{h}
     +  D^2 s^* \frac{n^2}{n-1} \frac{1}{h} \right]
\end{eqnarray}

The minimum of this equation is given by
\begin{equation} 
 \label{eqf:CPBS}
  2 h \dot h = \frac{a\gamma}{\eta} H^2_n , \quad
  h(t)=\sqrt{\frac{a \gamma}{\eta}} H_n t^{1/2}.    
\end{equation}
with
\begin{equation}
  \label{eqf:Hn}
  H_n = \left[ \frac{\alpha_1 \sqrt{s^*} }{\alpha_2} \frac{3(3n+2)(n+1)}{(n+2)(n^2+3n+3)} \right]^{1/2} .
\end{equation}
The tip of the finger advances with the Lucas-Washburn scaling $h \sim t^{1/2}$. 
Different profile predicts a different front factor $H_n$ for the tip position.

%===========================================================
\begin{figure}[htbp!]
  \includegraphics[width=0.6\columnwidth]{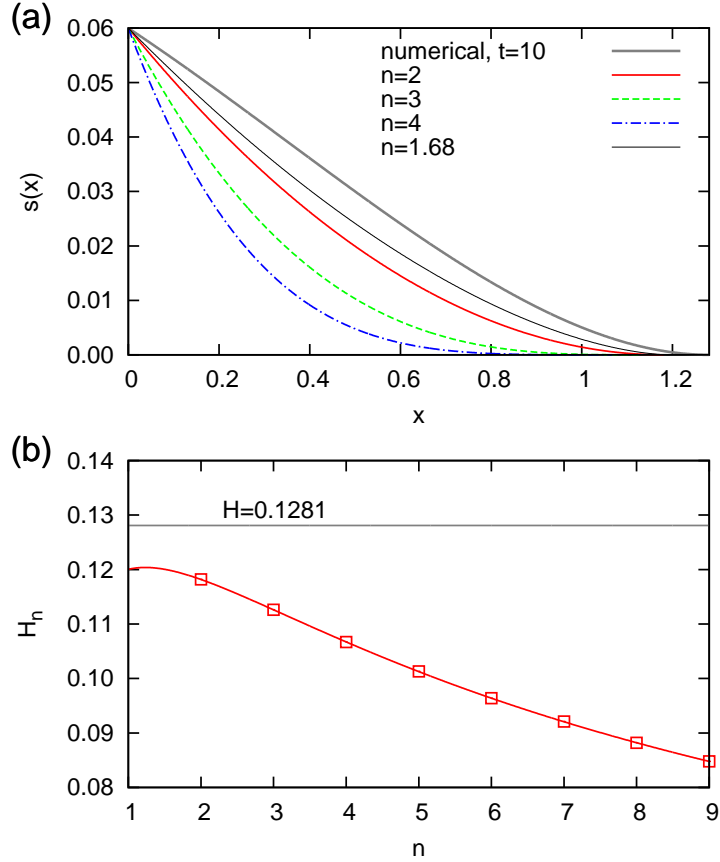}
  \caption{Comparison between the numerical solution to Eq.~(\ref{eqf:pde}) and approximated solutions Eq.~(\ref{eqf:CPBS}) for $n=2,3,4$ and $n=1.68$. (a) The saturation profiles are shown for $t=100/(a\eta/\gamma)$. (b) The front factor $H_n$ in the expression of the tip position. }
  \label{fig:fingerflow}
\end{figure}
%===========================================================

Figure \ref{fig:fingerflow} shows the comparison of the approximate solution $s(x,t)$ calculated by Eq.~(\ref{eqf:CPBS}) and the numerical solutions to Eq.~(\ref{eqf:pde}). 
We see that the saturation profile of $n=2$ is the closest to the numerical solution [Fig.~\ref{fig:fingerflow}(a)]. 
We can also compare the front factor $H_n$ in the tip position of the finger. 
They are given by Eq.~(\ref{eqf:Hn}) and shown in Fig.~\ref{fig:fingerflow}(b).
Numerical solution to the partial differential equation (\ref{eqf:pde}) with boundary condition (\ref{eqf:bc}) and initial condition (\ref{eqf:ic}) gives $H \simeq 0.1281$ \cite{Dong1995}.
Again, the graph indicates that the best estimate for $h(t)$ is obtained for $n=2$.

We can confirm this observation by calculating the minimum value of the modified Rayleighian (\ref{eqf:mRay})
\begin{equation}
  \label{eqf:mRaymin}
  \mRay = \frac{ \alpha_1^{3/2} (s^*)^{3/4} }{ \alpha_2^{1/2} }
            \frac{ (a\gamma)^{3/2} }{ \eta^{1/2} } C_n {t}^{-1/2} 
\end{equation}
with
\begin{eqnarray}
  C_n &=& -\frac{\sqrt{3}}{8} \frac{ (3n+2)^{3/2} }{ (n+1)^{1/2} (n+2)^{3/2} (n^2+3n+3)^{1/2}} \nonumber \\
  \label{eqf:Cn}
  && + \frac{1}{32\sqrt{3}}  \frac{n^2 (n+2)^{1/2} (n^2+3n+3)^{1/2} }{(n-1) (3n+2)^{1/2} (n+1)^{1/2}} 
\end{eqnarray}

\begin{figure}[!htbp]
  \includegraphics[width=0.6\columnwidth]{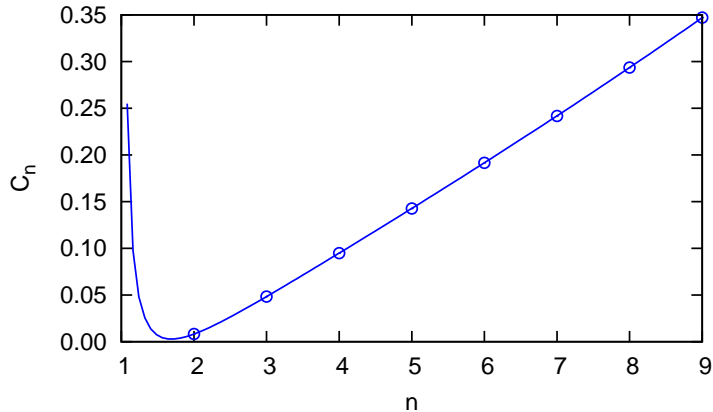}
  \caption{Comparison of different trail functions based on the Onsager Machlup integral. Here the value of $C_n$ from Eq.~(\ref{eqf:Cn}) is plotted as a function of $n$.}
  \label{fig:mRay}
\end{figure}

Fig.~\ref{fig:mRay} shows the value of $C_n$, which is proportional to the Onsager Machlup integral, as a function of $n$.  
It is seen that $C_n$ takes minimum at integer number $n=2$, in agreement with the results shown in Fig.~\ref{fig:fingerflow}. 
If we relax and allow $n$ to be a real number, we can see that $C_n$ has a minimum around $n\simeq 1.68$. 
The saturation profile for $n=1.68$ is also shown in Fig.~\ref{fig:fingerflow}(a) in black, which resembles the numerical solution even better than $n=2$.
This also demonstrates that the best approximation can be obtained by evaluating the Onsager Machlup integral.

%===========================================================
%===========================================================

% discussion

%%%%%%%%%%%%%%%%%%%%%%%%%%%%%%%%%%%%%%%%%%%%%%%%%%%%%%	
\section{Film Coating \label{sec:coating}}

\begin{figure}[htbp]
\includegraphics[width=0.6\columnwidth]{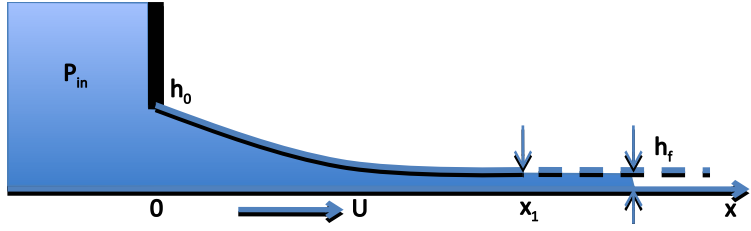}
\caption{Sketch of the coating system under consideration: 
a substrate is moving out at velocity $U$ from a liquid reservoir through a gap $h_0$
and is coated by the liquid.  The liquid pressure in the reservoir is kept at $P_{\rm in}$. }
\label{coating_illustrate}
\end{figure}

As the last example, we consider the problem of liquid coating on a solid substrate
(see Fig.~\ref{coating_illustrate}). A substrate is moving out from a
reservoir of viscous liquid with velocity $U$ through a gap $h_0$, and is 
coated with the liquid. The liquid in the reservoir is kept at a pressure $P_{\rm in}$ higher than 
the atmospheric pressure. (Here we consider the situation that $P_{\rm in}$ is negative, 
so the liquid is sucked in to the reservoir.)  We also assume 
that the gap $h_0$ is much smaller than the capillary length and ignore the effect of gravity.

We focus on the steady state of the system, and ask what is the thickness $h_f$ of the liquid 
film formed on the substrate in the steady state.  This question is slightly different 
from the previous ones.  In the previous examples, we discussed the time evolution of
non-equilibrium systems, but here we discuss the steady state of non-equilibrium systems.
The purpose of this example is to show that we can calculate the steady state 
directly, without solving the time evolution equation, using Onsager Machlup integral.

%===========================================================
%===========================================================
	
\subsection{Time evolution equation  for the film profile }
First we consider the time evolution of the film profile. 
We take $x$ coordinate parallel to the substrate, the origin of which is
located at the exit of the reservoir. Let $h(x,t)$ be the thickness of the 
liquid film at point $x$ and time $t$. 
The free energy of the system is given by the sum of the surface energy and the 
potential energy due the reservoir pressure $P_{\rm in}$
\begin{eqnarray} 
 A 
   &=& \gamma\int_0^{\infty}\left[ \sqrt{1 +\left(\frac{\partial h}{\partial x}\right)^2} -1 \right ]dx - P_{\rm in}\int_0^{\infty} h dx \nonumber \\
  \label{eqn:coat1}
   &=& \frac{\gamma}{2}\int_0^{\infty}\left(\frac{\partial h}{\partial x}\right)^2 dx - P_{\rm in}\int_0^{\infty} h dx, 
	\end{eqnarray}
where $\gamma$ is the surface tension, and 
we have assumed $(\partial h/\partial x)^2 \ll 1$.  

Let $v(x, t)$ be the depth averaged velocity of the fluid. 
The conservation condition of the fluid is written as
\begin{equation}\label{conservation}
	\dot{h} = -\frac{\partial}{\partial x}(hv).
\end{equation} 
$h(x,t)$ should satisfy the boundary condition 
	\begin{equation}\label{bc1}
	h(0,t) = h_0.
	\end{equation}
The free energy change rate $\dot A$ is calculated from Eqs.~(\ref{eqn:coat1}) and (\ref{conservation})
\begin{eqnarray}
	\dot{A} &=& \gamma\int_0^{\infty} \frac{\partial h}{\partial x} \frac{\partial\dot{h}}{\partial x} dx
	- P_{\rm in}\int_0^{\infty} \dot{h}dx\nonumber\\
	&=& \left [  \gamma\frac{\partial^2 h}{\partial x^2} hv +  P_{\rm in}hv
                     \right]_{x=0}   
                	- \gamma\int_0^{\infty} \frac{\partial^3 h}{\partial x^3} hv dx. 
\end{eqnarray}
where we have used that $\dot h$ is equal to 0 at $x=0$.
On the other hand, the energy dissipation function is calculated by the lubrication
approximation \cite{DiYana2018} as
\begin{equation}
	\Phi= \int_0^{\infty} \frac{3\eta}{2h} (v - U)^2  dx,
\end{equation}
where $\eta$ is the viscosity of the fluid. 
The velocity which minimizes the Rayleighian $\Phi + \dot A$ is given by 
\begin{equation}\label{velocity}
	v^* = U + \frac{\gamma}{3\eta}h^2\frac{\partial^3 h}{\partial x^3},
\end{equation}
The minimization of the Rayleighian also gives the following boundary condition:
\begin{equation}\label{bc2}
	\frac{\partial^2 h}{\partial x^2}\bigg|_{x=0}= \frac{- P_{\rm in}}{\gamma}.
\end{equation}

Substituting the velocity (\ref{velocity}) into the conservation law 
(\ref{conservation}), we have the evolution equation for the film thickness 
\begin{equation}\label{evolution}
  \dot{h} = -\frac{\partial}{\partial x}
     \left[  
        \frac{h^3}{3} \left(    \frac{\partial^3 h}{\partial x^3}    \right)  +\text{Ca}\, h 
      \right],
	\end{equation}
where $\text{Ca}= U/U^*$ is the capillary number and $U^*= \gamma /\eta $ is the capillary
velocity.

\subsection{Variational calculus for steady state using Onsager Machlup integral }
We now use the variational principle to  seek an approximate solution for the
steady state of Eq.~(\ref{evolution}).  The variational principle says that among all 
possible kinetic paths allowed for $h(x;t)$, nature chooses the path 
which minimizes the Onsager Machlup integral (\ref{eq:OM_integral1}). 

In the present problem, the the modified Rayleighian is given by
\begin{equation}  \label{eqn:coat2}
      \mRay = \int_0^{\infty} \frac{3\eta}{2h} (v - v^*)^2  dx,
\end{equation}
where we have again used the formula of Eq.~(\ref{eqn:15}).  In Eq.~(\ref{eqn:coat2}), 
$v$ is the velocity determined by the conservation equation (\ref{conservation}), and
$v^*$ is the velocity given by Eq.~(\ref{velocity}).  

In the steady state, $v$ and $v^*$ become independent of time.  
Therefore the minimization principle of the Onsager Machlup integral
is equivalent to the minimization of the modified Rayleighian $\mRay$.  

To seek the minimum of Eq.~(\ref{eqn:coat2}), 
we take the same strategy as in the previous examples; we consider certain 
form for $h(x)$ which includes certain parameters, and seek the minimum in the 
parameter space.  A simple choice for the liquid profile $h(x)$ is
\begin{equation}\label{coating_profile1}
   h(x) = 
	\left\lbrace 
	\begin{array}{lr}
	\frac{1}{2} a x^2 + b x + h_0, \quad & 0<x<x_f\\
	 h_f, & x>x_f
	\end{array}\right.
\end{equation}
This function is chosen to satisfy the condition $h(0)=h_0$ and $h(\infty)=h_f$.
Equation~(\ref{coating_profile1}) includes 4 parameters $a,b, x_f$ and $h_f$. 
The number of parameters is reduced to 1 
since $h(x)$ has to satisfy the boundary condition (\ref{bc2}), 
and the continuity condition for  $h(x)$ and $h'(x)$ at $x=x_f$.
However, Eq.~(\ref{coating_profile1}) has a problem that 
the second order differential $h''(x)$ is discontinuous at $x=x_f$, and 
therefore $v^*$ has a singularity of delta function type $\delta(x-x_f)$ which makes
the modified Rayleighian $\mRay[h(x)] $ diverge. 
Therefore, we considered the following form
\begin{equation}\label{coating_profile}
   h(x,t) = 
	\left\lbrace 
	\begin{array}{lr}
	\frac{1}{2} a x^2 + b x + h_0, & 0<x<x_1\\
	(h_1 - h_f) e^{-\kappa(x - x_1)} + h_f, & x>x_1
	\end{array}\right.
\end{equation}
This function includes 6 parameters, $a, b, x_1, h_1, h_f$ and ,$\kappa$, but the number of 
independent parameters can be reduced to 2 if we use the boundary condition
(\ref{bc2}), and the 3 continuity conditions for $h$, $h'$ and $h''$ at $x=x_1$.  
We have chosen $h_1$ and $h_f$ as independent parameters. The other parameters
are expressed by $h_1$ and $h_f$ as follows.
\begin{equation*}
	%\left\lbrace 
	\begin{array}{lcl}
	a &=& -\frac{P_{\rm in}}{\gamma}\\
	b &=& - \sqrt{a(2 h_0 - h_1 - h_f)} \\
	\kappa &=&\sqrt{\frac{a}{h_1 - h_f}} \\
	x_1 &=& \sqrt{\frac{2 h_0 - h_1 - h_f}{a}} - \sqrt{\frac{h_1 - h_f}{a}}
	\end{array} %\right.
\end{equation*}

At steady state, the flux $hv$ is constant and is given by $h_f U$. Hence 
the velocity $v(x)$ is given by
\begin{equation} \label{velocity1}
    v(x) = \frac{h_f U}{h(x)} .
\end{equation}
On the other hand, the velocity $v^*(x,t)$ is given by Eq.~(\ref{velocity}).

\begin{figure}[htbp]
  \includegraphics[width=1.0\textwidth]{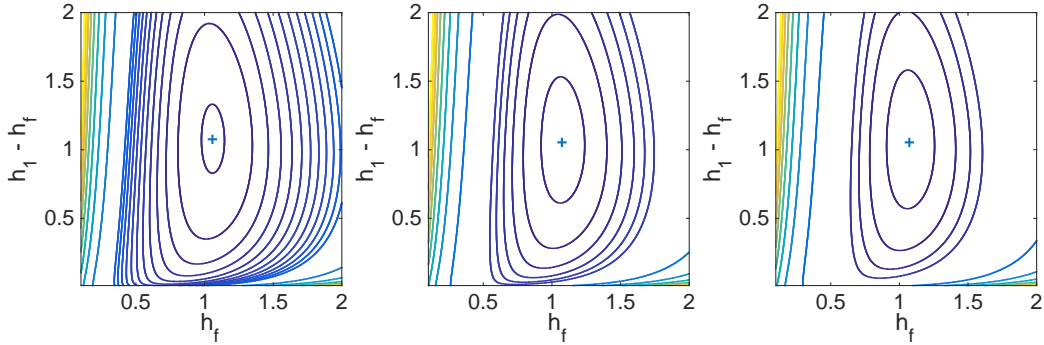}
\caption{The contour maps of $\mRay$ for $\tilde{h}_0=10, 100, 1000$, respectively. 
The $x$-axis represents $\tilde{h}_f$ and the $y$-axis represents $\tilde{h}_1-\tilde{h}_f$.}
		\label{coating_contour}
	\end{figure}

Using Eqs.~(\ref{velocity1}) and (\ref{velocity}), we calculated the modified Rayleighian 
for the profile (\ref{coating_profile}), and obtained the following result
\begin{multline}\label{coating_ray}
	\mRay(h_1, h_f) 
	= \frac{3\eta U^2}{2}\left(
	\frac{2h_1^2 + h_f^2}{\sqrt{a}(h_1 + h_f)^{5/2}} \left(\arctan\sqrt{\frac{2h_0}{h_1 + h_f} - 1} - \arctan\sqrt{\frac{h_1 - h_f}{h_1 + h_f}}\right)\right.\\
	+ \frac{h_f}{2\sqrt{a}(h_1 + h_f)^2}\left(\left(4 - \frac{h_f^2}{h_1^2}\right)\sqrt{h_1 - h_f} +
	\left(\frac{h_1h_f + h_f^2}{h_0^2} - \frac{4h_1 + h_f}{h_0}\right)\sqrt{2h_0 - h_1 - h_f}\right)\\
	\left.+ \frac{(h_1 - h_f)^{5/2}}{2\sqrt{a} h_1^2 h_f} - \frac{a(h_1 - h_f)}{3\text{Ca}} + 
	\frac{a^{5/2}}{9\text{Ca}^2\sqrt{h_1 - h_f}}\left(\frac{1}{5}h_1^3 + \frac{3}{20}h_fh_1^2 + \frac{1}{10}h_f^2h_1 + \frac{1}{20}h_f^3\right)
	\right).
\end{multline}
Using dimensionless parameters $\tilde{h}_0 = ah_0/\text{Ca}^{2/3}$,
 $\tilde{h}_1 = ah_1/\text{Ca}^{2/3}$, $\tilde{h}_f = ah_f/\text{Ca}^{2/3}$, 
the above equation can be written as
\begin{multline} \label{eqn:coat10}
	\mRay(\tilde{h}_1, \tilde{h}_f) = \frac{3\eta U^2}{2}\text{Ca}^{-1/3}\left(
	\frac{2\tilde{h}_1^2 + \tilde{h}_f^2}{(\tilde{h}_1 + \tilde{h}_f)^{5/2}} 
	\left(\arctan\sqrt{\frac{2\tilde{h}_0}{\tilde{h}_1 + \tilde{h}_f} - 1} - \arctan\sqrt{\frac{\tilde{h}_1 - \tilde{h}_f}{\tilde{h}_1 + \tilde{h}_f}}\right)\right.\\
	\frac{\tilde{h}_f}{2(\tilde{h}_1 + \tilde{h}_f)^2}\left(\left(4 - \frac{\tilde{h}_f^2}{\tilde{h}_1^2}\right)\sqrt{\tilde{h}_1 - \tilde{h}_f}+
	\left(\frac{\tilde{h}_1\tilde{h}_f + \tilde{h}_f^2}{\tilde{h}_0^2} - \frac{4\tilde{h}_1 + \tilde{h}_f}{\tilde{h}_0}\right)\sqrt{2\tilde{h}_0 - \tilde{h}_1 - \tilde{h}_f}\right)\\
	\left.+ \frac{(\tilde{h}_1 - \tilde{h}_f)^{5/2}}{2\tilde{h}_1^2 \tilde{h}_f} - \frac{\tilde{h}_1 - \tilde{h}_f}{3} + 
	\frac{1}{9\sqrt{\tilde{h}_1 - \tilde{h}_f}}\left(\frac{1}{5}\tilde{h}_1^3 + \frac{3}{20}\tilde{h}_f\tilde{h}_1^2 + \frac{1}{10}\tilde{h}_f^2\tilde{h}_1 + \frac{1}{20}\tilde{h}_f^3\right)
	\right).
\end{multline}
Minimization of Eq.~(\ref{eqn:coat10}) with respect to $\tilde{h}_1$ and $ \tilde{h}_f$ gives the
steady state film thickness $h_f$.  Equation (\ref{eqn:coat10}) indicates that 
$ \tilde{h}_f$ depends on $\tilde{h}_0$, but does not depend on Ca.  
Figure~\ref{coating_contour} shows the contour maps of $\mRay$ for various values of $\tilde{h}_0$, 
$\tilde{h}_0=10, 100, 1000$.
The $x$-axis represents $\tilde{h}_f$ and the $y$-axis represents $\tilde{h}_1-\tilde{h}_f$.

%===========================================================
%===========================================================

\begin{figure}[htbp]
  \includegraphics[width=0.7\textwidth]{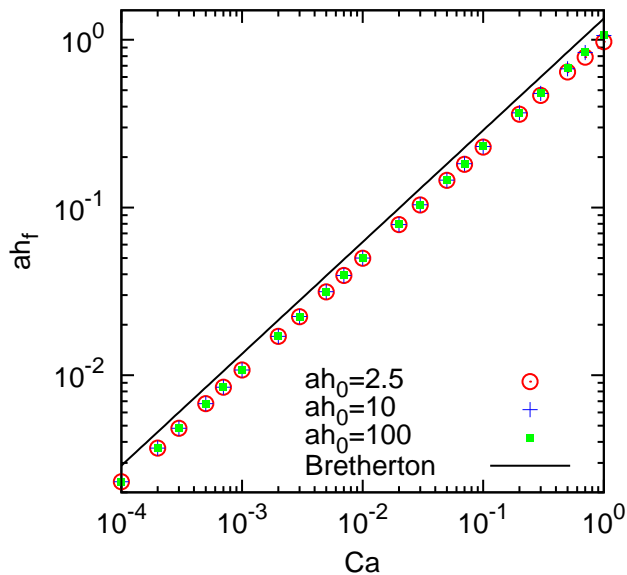}
  \caption{Film thickness $ah_f$ as a function of Ca. The symbols represents the minimum of $\mRay$ by numerical calculation. 
%The blue symbols represents the numerical results of the PDE (\ref{evolution}). 
The black solid line represents Bretherton's $2/3$ power law \cite{Bretherton}. }
\label{coating_thickness}
\end{figure}

Figure~\ref{coating_contour} shows that the minimum position is almost independent of
of $\tilde{h}_0$, indicating that the film thickness is given by
$\tilde{h}_f(\approx1.07)$ and is almost independent of $\tilde{h}_0$，
	which gives us the celebrated scaling law
\begin{equation}\label{bretherton}
	a h_f \approx 1.07 \left( \textrm{Ca} \right)^{2/3}.
	\end{equation}
This can be compared with result of the asymptotic solution  
for the steady state for Eq.~(\ref{evolution}) \cite{Bretherton, Carvalho2000}, which gives 
the same scaling law as Eq.~(\ref{bretherton}), but the numerical coefficient
is different, equal to 1.34.  
Figure~\ref{coating_thickness} shows the comparison 
between the asymptotic analysis and the variational calculus.  Our 
calculation indicates that the effect of gap distance $h_0$ is small.

%===========================================================
%===========================================================

\section{Conclusion}
In this paper, we have shown a new usage of Onsager Machlup variational principle.  
It is based on the minimum principle that nature chooses the kinetic path which makes the Onsager Machlup integral $\mathcal{O}[x(t)]$ minimum.  
We have demonstrated that this principle can be used to get approximate solutions for the evolution equations for linear and non-linear diffusion equations. 
\change{These approximate solutions in general involve one or a few time-dependent parameters, such as $a(t)$ in the diffusion problem (\ref{eqd:trial}) and $h(t)$ in the finger flow problem (\ref{eqf:on}). 
These parameters are optimized step by step in time, determined by the kinetic equations derived from Onsager principle. 
On the other hand, there are also one or a few parameters that remain constant in time, such as the power $m$ in Eq. (\ref{eqd:trial}) and $n$ in Eq. (\ref{eqf:on}). 
These time-independent parameters represent the functional form of the approximate solutions. 
They cannot be optimized locally in time, yet they can still be optimized globally over a period of time through the minimization of Onsager Machlup integral.
We have shown that the Onsager Machlup integral can be used to obtain the steady state 
without solving the evolution equation in our coating example.
}

\change{
The present method may be regarded as a special kind of least square minimization method, but there is a significant difference. 
In the existing least square minimization methods, the choice of the evaluation function (the function which tells the goodness of the solution) is arbitrary.  
If the evolutions equations is given by a set of equations,  one can put any weight for each equation to get the evaluation function, and the final result depends on the choice of the weight.  
What we are proposing in this paper is that the Onsager Machlup integral is the most natural choice for the evaluation function since it represents the fundamental quantity which governs the physics in the problem.  
}

Many extensions are possible based on the Onsager Machlup integral.  
The method may be used
to find new numerical scheme to solve the kinetic equations.  It can also 
be used to obtain oscillatory solutions in space and time for non-linear 
partial differential equations. More examples will be shown in future which 
demonstrate the powerfullness of Onsager's variational principle.

%%%%%%%%%%%%%%%%%%%%%%%%%%%%%%%%%%%%%%%%%%%%%%%%%%%%%%
\begin{acknowledgments}
M.D. thanks Prof. Hans Christian {\"O}ttinger of ETH  for his illuminating
discussion on the Onsager principle which initiated this work. 
This work was supported by the National Natural Science Foundation of China (NSFC) through the Grant No. 21504004 and 21774004. 
M.D. acknowledges the financial support of the Chinese Central Government in the Thousand Talents Program. 
\end{acknowledgments}

%%%%%%%%%%%%%%%%%%%%%%%%%%%%%%%%%%%%%%%%%%%%%%%%%%%%%%%%%%%%%%%%%%%
\bibliography{Onsager}

% uncomment the following lines if using preprint endfloats
%\listoffigures

\end{document}